# A Human-Centered Approach to Data Privacy : Political Economy, Power, and Collective Data Subjects

**Meg Young**
University of Washington
Seattle, WA 98105, USA
megyoung@uw.edu

## Abstract
Researchers find weaknesses in current strategies for protecting privacy in large datasets: many 'anonymized' datasets are re-identifiable, and norms for offering data subjects notice and consent over-emphasize individual responsibility. Based on fieldwork with data managers in the City of Seattle, I identify ways that these conventional approaches break down in practice. Drawing on work from theorists in sociocultural anthropology, I propose that a Human-Centered Data Science move beyond concepts like dataset identifiability and sensitivity toward a broader ontology of who is implicated by a dataset, and new ways of anticipating how data can be combined and used.

## ACM Classification Keywords
H.5.m. Information interfaces and presentation (e.g., HCI): Miscellaneous.

**Introduction**

Data science researchers commonly navigate a particular tension between an ethical mandate to protect their participants' privacy, and, where necessary, to preserve the richness of granular datasets. Conventional approaches to managing privacy include (1) de-identification, (2) notice and consent, and (3) collection limitation, and (4) obfuscation. In this paper, I draw on fieldwork with the City of Seattle to point to ways that these approaches cannot protect individual privacy from threats like re-identification and inferential analysis. Then, I suggest new directions for addressing data privacy.

This paper draws on 6 months of fieldwork, including 24 interviews across seven departments in the City of Seattle between January-May 2015, we explored the ways that municipal data managers consider privacy in their collection, management, and release of data [1]. These questions are especially salient in the Washington state context, where the WA Public Records Act (Ch. 42.56 RCW) provides for all public records to become available in response to a public disclosure request, subject to few limitations. In this paper, I develop insights from our work on municipal data into considerations for researchers in a Human-Centered Data Science.

Datasets can be more or less identifiable to a particular *data subject*, an individual contained within the dataset. They can also be more or less sensitive, as defined by a potential for privacy harm, depending on the nature of the data [2]. As a starting point for this paper, I contend that the conceptual schema of 'identifiability' and 'sensitivity' no longer serve us, given the changing realities about how data can be re-identified or used to make inferences about groups.

Data de-identification is defined by the removal of personally identifiable information (PII), for example, name and address fields. Seemingly innocuous de-identified records can be tied together in an operation called a 'join;' an operation for combining datasets by connecting their common attributes. Seminal work by Latanya Sweeney and others has demonstrated the removal of PII method is still vulnerable to re-identification by correlating details from a dataset with publicly available, identified data [3]. Ohm, Felten and Naranayan state that there is no evidence that de-identification works to protect privacy or anonymity in practice [4] [5]. In our research, we conducted joins across datasets on Seattle's already-open municipal data, in order to demonstrate that this was true of open municipal data as well.

A second, widely-held approach to preserving privacy is to follow principles set forth by the Federal Trade Commission, called the Fair Information & Privacy Practices (FIPPS, or metonymically, 'notice and consent'). Under FIPPS guidelines, data subjects must be notified at the time that data is gathered about them, briefed about the uses to which data will be put, and asked to consent. FIPPS provide data subjects the ability to opt-in or opt out of inclusion in a dataset. Perhaps because of the difficulty of following FIPPS in practice, it is sometimes raised as a 'gold standard' for maintaining data subject privacy [6]. However, scholars have also raised doubts about the extent to which 'notice and consent' can protect privacy in a meaningful way, because it places too much responsibility on the user's ability to decide. It does not account for the staggering number of requests that would require consent, or the users' cognitive overload that would result [7]. In our research with the City of Seattle, we recommended that municipal data follow 'notice and consent' guidelines. However, this was perceived as infeasible for the scale and complexity of the city's data practices.

One of the FIPPs, collection limitation, is itself a strategy to protect privacy. In our work in Seattle, we observed value tensions around collection limitations. Rich, detailed data was valuable for reasonable purposes, for example, studies on traffic volume and commute time in the Department of Transportation. Where high-resolution data like location traces were a liability, the city depended on private contractors to handle the raw data and return aggregated statistics. In practice, one strategy we observed was to 'collect first, aggregate later,' such that very sensitive data was being collected, and then distilled or discarded as the data circulated between the sensor, vendor, and city departments. Thus, collection limitation by the city does not always result in collection limitation by all relevant actors. This presents a privacy risk in that any data that is collected can be re-circulated or spilled.

Researchers have proposed other methods for protecting data. Foremost among these is the differential privacy approach: instead of having access to all disaggregated records, differential privacy requires queries to a protected database [8]. Results from the queries use sophisticated statistical techniques to shield individuals within the dataset, by producing statistically consistent noise [8]. Whereas differential privacy offers an approach for institutional adoption, Nissenbaum and Brunton call on us as individuals to deliberately obfuscate our own data trails, as a weapon to counter asymmetry in the ways their data can be stored and used [9].

In the following prompts, I offer several directions for a Human-Centered Data Science to move beyond these conventional approaches as a means for protecting data privacy.

*1. Hacks, obfuscation, and 'folk data practices'*
In her seminal essay, 'Can the subaltern speak?' Gayatri Spivak draws attention to the way that members of relatively oppressed groups are homogenized into a single category and 'spoken for' [10]. Inspired by Spivak, a Human-Centered Data Science might interrogate the way our categories as researchers are constructed. We may scaffold our practice with critical questions about who is included in the data and who isn't. How can we design datasets that preserve multiplicities of experience, instead of flattening them along a set of spectra? Inspired by Spivak's vision, a Human-Centered Data Science may begin to open up inquiries into 'folk data practices,' or trace ways that data subjects thwart attempts to have data collected about them. How does a notion like obtaining user 'consent' assume that users are empowered to give it, or that it will be heard in the form in which they express (or withhold) it?

*2. From individual to collective data subjects*
As Cate notes, conventional approaches to preserving online privacy puts the burden on individuals and their choices [11]. Similarly, in large datasets, we might ask how the idea of protecting privacy is focused around the idea of individual privacy. Real-world examples point to ways that companies have leveraged data about users in neighborhoods [12], racial groups [13], and type of hardware [14]. By thinking of privacy as applying to places, identities and things, we can begin to break down received notions of privacy violations as only occurring on an individual level. A Human-Centered Data Science may develop ways to anticipate these other scales of impact, to include race and social justice implications of datasets.

### 3. The political economy of datasets

In our research with the City of Seattle, we found that leaders of the open data initiative wanted to realize latent value in the data for local community members and local enterprise. Conspicuously absent from their concept of who uses open data was the data brokerage industry. Data brokers aggregate, index, and sell data to advertisers; a recent FTC report found that the majority of data brokers studied relied on open government data [15]. A Human-Centered Data Science will create venues to discuss the markets that datasets enter as they are collected, managed and released. It will offer computational social scientists ways to anticipate unintended users of large-scale data, and ask questions about who benefits from it, and who is liable for it.

### 4. Disciplinary power and user exposure

Our current approaches to privacy can sometimes flatten our ethical vision—somewhat like digital image compression making an image blurry and low-resolution. For example, under FIPPS, data accuracy is an ethical standard; FIPPS encourages data holders to create means by which data subjects can correct and update their entries. The hope is that inaccurate data will not be used to data subjects' detriment, as in the prospect of being denied credit. However, while inaccurate data may harm users, so too may accurate data. James C. Scott argues that accurate data, like a census, makes data subjects legible and amenable to control [16]. This example shows us that considering ethics within a dataset may not always boil down to whether data is accurate, or whether data subjects may update their information. We may instead ask questions about what activities a dataset makes possible. By closing the loop, and bringing the answers to these questions back to the data collection stage, a Human-Centered Data Science can offer researchers a set of tools for anticipating and remediating ethical issues latent in a research dataset.

### 5. A typology of open data

When data is made 'open,' how is it opened? On what platforms is it hosted, and in what formats is it made available? Who can read these formats, and on what machines? The word 'open' obscures the answers to these questions, by presenting multiple forms of openness as a single heading, and valorizing it. In our fieldwork with the City of Seattle, we observed mounting pressures behind opening data for government transparency, from the enforcement of public records laws to campaigns by activists. The local community using the data, however, found that geospatial data was often released in expensive proprietary formats, making it unusable to them. A Human-Centered Data Science would be an appropriate community for scholarship that interrogates values behind open data initiatives. It may also offer new ways of categorizing openness, so that researchers working with large datasets might no longer refer to 'open' data as an undifferentiated whole, but as a set of choices with respect to platforms, formats, licensing, and uses.

Based on fieldwork in Seattle, and drawing on sociocultural anthropological theory, I have suggested that that Human-Centered Data Science can go beyond existing approaches to protecting individual privacy in large-datasets in the following ways: (1) attend to hacks, obfuscation and 'folk' data practices, (2) move from individual to collective data subjects, (3) analyze the political economy of datasets, (4) examine disciplinary power and user exposure, and (5) create a typology of open data.